\documentclass[prl,twocolumn]{revtex4}
\usepackage{graphicx}%
\usepackage{dcolumn}
\usepackage{amsmath}
\begin{document}

\preprint{HEP/123-qed}

\title{Giant Low-temperature Piezoresistance Effect in AlAs Two-dimensional Electrons}
\author{Y. P. Shkolnikov, K. Vakili, E. P. De Poortere, and M. Shayegan}
\affiliation{Department of Electrical Engineering, Princeton
University, Princeton, New Jersey 08544}

\date{\today}

\begin{abstract}
We present piezoresistance measurements in modulation doped AlAs quantum wells
where the two-dimensional electron system occupies two conduction band valleys
with elliptical Fermi contours. Our data demonstrate that, at low temperatures,
the strain gauge factor (the fractional change in resistance divided by the
sample's fractional length change) in this system exceeds 10,000. Moreover, in
the presence of a moderate magnetic field perpendicular to the plane of the
two-dimensional system, gauge factors up to 56,000 can be achieved.  The
piezoresistance data can be explained qualitatively by a simple model that
takes into account intervalley charge transfer.
\end{abstract}


\maketitle

The piezoresistance effect \cite{Bridgman1932} -- the change of a device's
electrical resistance as a function of applied stress -- is of great
technological interest. It is used to make force, displacement, and pressure
sensors \cite{Marian1992} that are ubiquitous in various industries.  Even the
semiconductor industry is using the piezoresistance effect to improve the
performance of field-effect transistors \cite{Keyes2002}. Recently,
piezoresistivity has found use in measuring sub-nanometer ($\sim$0.01nm)
displacements of tips in atomic force microscopes
\cite{Tortonese1993,Harley2000}. With the advancements in science and
technology of materials and devices whose size approaches the atomic scale, the
interest in sensors that can detect ultra-small forces and distances is likely
to grow even higher. Piezoresistance is also useful as a probe of the
electronic properties of materials.

There are two main sources of the piezoresistance of a strain gauge. First,
applied stress changes the resistance of the gauge by altering the length and
the cross sectional area of the gauge. This geometrical distortion is the
primary source of piezoresistance in metallic films. The sensor's gauge factor,
defined as the fractional change in the film's resistance divided by the
fractional change in its length, is determined primarily by the solid's Poisson
ratio, and is typically about 2. Second, stress can modify the electronic
structure of a solid and lead to large changes in its resistivity. This change
can be particularly significant in indirect band gap semiconductors such as Si
and Ge (and AlAs), where the conduction electrons occupy multiple minima
(valleys) in the conduction band. In this case, uniaxial or shear stress breaks
the symmetry of the band structure, lifts the degeneracy of the valley
energies, and causes an intervalley charge transfer. Since the valleys often
have anisotropic effective masses, such a transfer leads to significant changes
in electron mobility, and therefore resistance, along certain crystallographic
directions of the solid. This phenomenon is well-known for bulk Si and Ge
\cite{Smith1954,Herring1955} as well as for two-dimensional electron systems
(2DESs) in Si-MOSFETs \cite{Eisele1978} where gauge factors as large as
$\sim1,500$ have been observed. Here we report a giant low-temperature
piezoresistance effect, characterized by gauge factors exceeding 10,000, for
2DESs confined to modulation-doped AlAs quantum wells. Moreover, we show that
even much larger gauge factors can be achieved if the sample is placed in a
moderate, perpendicular magnetic field.

AlAs is an indirect gap semiconductor with conduction-band valleys at the
Brillouin zone X-points. Its constant energy surface consists of three highly
anisotropic ellipsoids (six half-ellipsoids) with longitudinal ($m_l)$ and
transverse ($m_t$) effective masses $\simeq1.0$ and $0.2$ (in units of free
electron mass), respectively. We designate these valleys by the direction of
their major axes, which point along the $<$100$>$ directions. In AlAs quantum
wells wider than $\sim 5$ nm grown on GaAs (001) substrates, because of the
GaAs-AlAs lattice mismatch strain, only the two valleys with their major axes
lying in the plane of the 2DES are occupied \cite{Maezawa1992}. While in the
absence of an external stress these valleys are energy degenerate, a
compression along either $[$100$]$ or $[$010$]$ breaks this degeneracy and
causes a redistribution of electrons between the valleys. Although the band
structure of Si and AlAs are similar, the valleys in Si are centered around the
six $\Delta$-line points that are away from the Brillouin zone edge. As a
result, there are six valleys in Si, two valleys along each of the $<$100$>$
directions. An additional difference is that, in the absence of applied stress,
in (001) Si-MOSFETs two out-of-plane valleys are occupied, while in a wide
(001) AlAs quantum well the 2D electrons occupy two in-plane valleys.

Our samples contain 2D electrons confined to an 11 nm-wide, AlAs quantum well
grown on a GaAs (001) substrate by molecular beam epitaxy \cite{Etienne2002}.
The quantum well is flanked by Al$_{0.4}$Ga$_{0.6}$As barrier layers and is
modulation doped with Si. Using a metal electrode (gate) deposited on the top
surface, we can vary the 2DES density $n$, between $2.85\times10^{15}$ and
$7.25\times10^{15}$ m$^{-2}$. To assure a uniform strain in the 2D layer, the
substrate thickness is kept below 250 $\mu$m. The resistance is measured using
a lock-in amplifier at 0.3 K on an etched Hall bar mesa aligned with the [100]
crystal direction. Piezoresistance measurements were performed on three
samples. Here we concentrate on the data from one sample, with other samples
corroborating the results.

We strain the sample by gluing it to a commercial piezoelectric stack (piezo)
with the [100] crystal direction aligned to the poling direction of the piezo
\cite{shayegan2003}. To shield the 2DES from the electric fields in the stack,
we evaporated a 120 nm thick \ Ti/Au gate on the back of the sample's substrate
and kept it under constant bias throughout the measurements. Under a positive
bias applied to the piezo, its surface expands along the poling direction and
shrinks in the perpendicular direction. This deformation is transmitted to the
sample through the glue. Using a calibrated, metal strain gauge glued to the
piezo's opposite side, we monitor the applied strain with a relative accuracy
of 5\%.  Via this technique, we can vary the shear strain $\epsilon$
(difference in the fractional change of length along the [100] and [010]
directions) by $4.7\times10^{-4}$. Since the negative of the ratio of the
lateral to the longitudinal deformation of the piezo is 0.38
\cite{shayegan2003}, close to the Poisson ratio of AlAs ($r=0.32$), the
resulting sample stress is almost uniaxial; tensile stresses along [100] and
[010] are equal to 73$\epsilon$ and 8.3$\epsilon$ (in GPa), respectively. All
other stress tensor components are zero.

\begin{figure}
\includegraphics[scale=0.8]{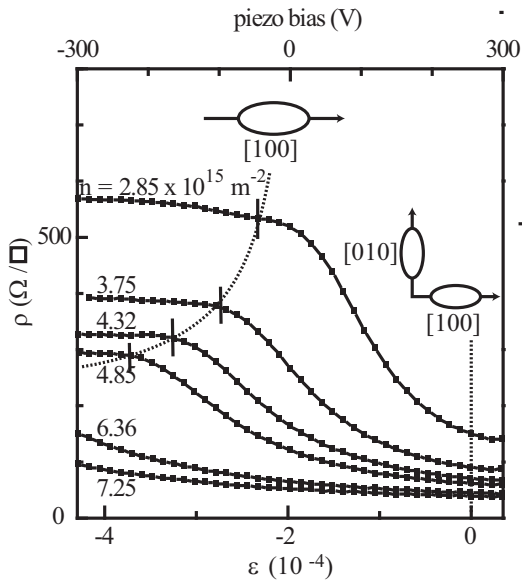}
\caption{Resistivity of an AlAs 2DES along the [100] direction vs.
strain for various densities as indicated. Compression along [100]
results in a transfer of the electrons from the [010] valley to
the [100] valley. Since the electron mobility along [100] is
smaller for the [100] valley, the resistivity of the system
increases with increasing compression. The valleys are equally
occupied at zero strain (right dotted line), while the left dotted
line indicates the strain at which the [010] valley depopulates.
Maximum strain gauge factor is $11,700$ for $n=2.85\times10^{15}$
m$^{-2}$. The sample is already strained at zero piezo bias
because of the anisotropic deformation of the piezo stack during
the cooling.}
\end{figure}

We show the results of our longitudinal piezoresistance measurements in Fig. 1.
Each trace corresponds to the resistivity $\rho$ along [100] at a constant 2DES
density, with the piezo bias swept from -300V to 300V. As the sample is
compressed along the [100] direction, the resistivity of the sample first
increases with increasing magnitude of strain, and then saturates. This
behavior of the resistivity is consistent with the transfer of electrons out of
the [010] valley (with smaller effective mass along [100]) and into the [100]
valley (with larger effective mass along [100]). The transfer of charges is
corroborated by the change of the frequency composition of the Shubnikov de
Haas oscillations. Using these oscillations, we determine and mark two strains
for each $\rho$ trace; right dotted line indicates the (zero) strain at which
both valleys are equally occupied, while the left dotted line indicates the
strain at which the [010] valley depopulates.

Since the resistivity in our sample is a nonlinear function of strain, to
quantify the piezoresistance we concentrate on the differential gauge factor
$k=(1+r)(d\rho/\rho)/(d\epsilon)$. The maximum $k$ value varies between 11,700
and 6,200 for the density range $n=(2.85 - 7.25)\times10^{15}$ m$^{-2}$. A
quantitative fit of the entire piezoresistivity curve requires a detailed
knowledge of both intervalley and intravalley scattering mechanisms
\cite{Keyes1960}. However, we can understand the major features of this curve
using a simple model in which we assume that the electron scattering lifetimes
in the two valleys are equal, isotropic, and independent of the electron
concentration in either valley, and ignore intervalley scattering. We also use
$E_2=5.8$ eV for the electron shear deformation potential of AlAs
\cite{Charbonneau1991}. This model predicts that for a strain of
$n\pi\hbar^2/E_2\sqrt{m_lm_t}$, which transfers all the electrons into a single
valley, the resistance increase should be $0.5(1+m_l/m_t)$. For $n=(2.85 -
4.85)\times10^{15}$ m$^{-2}$, the predicted strain, $(2.6-4.3)\times10^{-4}$,
and the factor of $\simeq3$ increase in resistance are consistent with the
experimentally observed values.

Our maximum strain gauge factor ($k$=11,700 for $n=2.85\times10^{15}$ m$^{-2}$
is about 8 times larger than previously reported in (001) Si-MOSFETs
\cite{Eisele1978}, although the deformation potential of Si (8.2 eV) and AlAs
are close as are the electron effective masses. A factor of two of this
increase can be attributed to the higher density ($n=1.1\times10^{16}$
m$^{-2}$) of the Si samples that were studied \cite{Eisele1978} and to the band
structure differences between Si and AlAs. We do not know what causes the
remaining difference, but speculate that it is the result of the non-monotonic
strain dependence of the resistance caused by the strong intervalley scattering
in Si samples \cite{Dorda1978}.

\begin{figure}
\includegraphics[scale=0.8]{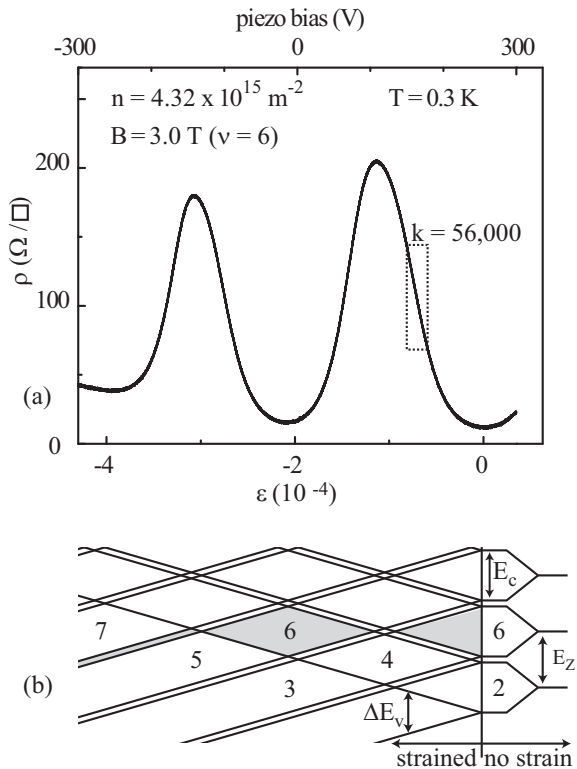}
\caption{(a) Piezoresitivity of the AlAs 2DES in the presence of a
3T perpendicular magnetic field (Landau level filling factor
$\nu=6$). Resistivity of the sample oscillates as the Landau
levels cross, closing and opening up the gap at the Fermi energy.
Gauge factor in the boxed region is $\sim$56,000. (b) Schematic
DOS diagram for the AlAs 2DES in the presence of magnetic field.
Electron energies are restricted to a discrete set of Landau
levels whose energy separation is controlled by the cyclotron
energy $E_c$, Zeeman splitting $E_Z$ between the oppositely
polarized spins, and the valley splitting $\Delta E_v$.}
\end{figure}

In the presence of a moderate magnetic field $B$ perpendicular to
the AlAs 2DES plane, the piezoresistivity increases dramatically,
with observed strain gauge factors as large as 56,000 [see Fig.
2(a)]. This giant piezoresistance effect results from the change
of the 2DES density of states (DOS) in $B$. With a finite $B$, the
orbital motion of electrons becomes quantized; the DOS [shown in
Fig. 2(b)] is no longer a constant function of energy, and instead
has non-zero values only at energies of the Landau levels. If the
magnitude of $B$ is such that an integer number of Landau levels
are filled (integer filling factor $\nu$), the system's Fermi
energy $E_F$ falls in an energy gap (DOS gap). This gap inhibits
electron scattering at low temperatures, leading to a smaller 2DES
resistivity. With applied stress, the Landau levels corresponding
to different valleys cross, causing the gap to close and the 2DES
resistivity to increase. Since the resistivity depends
exponentially on the size of the energy gap, its change can be
much larger than the effective mass anisotropy. The gap at $E_F$
vanishes when $\Delta E_v$ is equal to the cyclotron energy
($E_c$) of the system. Since $E_c<E_F$ (for a large range of $B$),
the Landau level crossings occur at strains smaller than required
to depopulate the [010] valley. The greater change in resistivity
over a smaller strain change results in strain gauge factors that
are much larger than those possible at zero $B$.

Nonlinearity of the piezoresistive response outside a narrow range of strain
limits the usefulness of the in-field strain gauge to applications in which
only small deformation is present. One can extend the measurable strain range
by employing a technique inspired by Michelson interferometry. As seen in Fig.
2(a), the resistance oscillates periodically with strain in a fixed magnetic
field (see also \cite{shayegan2003}). These periodic oscillations, which are
easy to detect thanks to the large ($\sim20$) ratio of maximum to minimum
$\rho$, allow one to accurately measure very large strains just by counting the
number of oscillations.

We have shown that a combination of confinement and magnetic field leads to a
giant piezoresistive response in AlAs 2DESs at low temperatures. This effect
may find important applications. Indeed, the gauge factor of 56,000 is large
enough to detect nanometer elongations of a centimeter-long sample by a simple
multimeter. With improved sample quality, the gauge factor can be further
increased at both zero and high magnetic fields. At $B=0$, one can achieve a
higher gauge factor by further lowering the density of the 2DES. At high $B$,
the 2DES can undergo a strain-induced transition from a quantum Hall state
(with nearly zero resistance) to a very high resistance neighboring
(insulating) state, with a very high resistance, leading to an even larger
piezoresistance effect. One-dimensional confinement, such as in a ballistic
quantum point contact \cite{vanHouten1988}, could also result in large gauge
factors; the resistance of the point contact can be extremely sensitive to the
number of occupied one-dimensional channels and therefore the valley occupancy.

We thank the ARO, NSF, and the Humboldt Foundation for support,
and K. Karrai for illuminating discussions.

\bibliography{L04-2855APL}

\end{document}